\documentclass[prd,twocolumn,a4paper,superscriptaddress]{revtex4}
\usepackage{graphicx,amssymb}
\begin{document}
\newcommand{\be}{\begin{equation}}
\newcommand{\ee}{\end{equation}}
\newcommand{\bq}{\begin{eqnarray}}
\newcommand{\eq}{\end{eqnarray}}
\newcommand{\bsq}{\begin{subequations}}
\newcommand{\esq}{\end{subequations}}
\newcommand{\bc}{\begin{center}}
\newcommand{\ec}{\end{center}}
\newcommand {\R}{{\mathcal R}}
\newcommand{\al}{\alpha}
\newcommand\lsim{\mathrel{\rlap{\lower4pt\hbox{\hskip1pt$\sim$}}\raise1pt\hbox{$<$}}}
\newcommand\gsim{\mathrel{\rlap{\lower4pt\hbox{\hskip1pt$\sim$}}\raise1pt\hbox{$>$}}}

\title{Clustering Properties of Dynamical Dark Energy Models}
\author{P. P. Avelino}
\email[Electronic address: ]{ppavelin@fc.up.pt}
\affiliation{Centro de F\'{\i}sica do Porto, Rua do Campo Alegre 687, 4169-007 Porto, Portugal}
\affiliation{Departamento de F\'{\i}sica da Faculdade de Ci\^encias
da Universidade do Porto, Rua do Campo Alegre 687, 4169-007 Porto, Portugal}
\author{L.M.G. Be{\c c}a}
\email{luis.beca@fc.up.pt}
\affiliation{Centro de F\'{\i}sica do Porto, Rua do Campo Alegre 687, 4169-007 Porto, Portugal}
\affiliation{Departamento de F\'{\i}sica da Faculdade de Ci\^encias
da Universidade do Porto, Rua do Campo Alegre 687, 4169-007 Porto, Portugal}
\author{C.J.A.P. Martins}
\email[Electronic address: ]{Carlos.Martins@astro.up.pt}
\affiliation{Centro de Astrof\'{\i}sica, Universidade do Porto, Rua das Estrelas s/n, 4150-762 Porto, Portugal}
\affiliation{DAMTP, University of Cambridge, Wilberforce Road, Cambridge CB3 0WA, United Kingdom}

\date{1 February 2008}
\begin{abstract}
We provide a generic but physically clear discussion of the clustering properties of dark energy models. We explicitly show that in quintessence-type models the dark energy fluctuations, on scales smaller than the Hubble radius, are of the order of the perturbations to the Newtonian gravitational potential, hence necessarily small on cosmological scales. Moreover, comparable fluctuations are associated with different gauge choices. We also demonstrate that the often used homogeneous approximation is unrealistic, and that the so-called dark energy mutation is a trivial artifact of an effective, single fluid description. Finally, we discuss the particular case where the dark energy fluid is coupled to dark matter.
\end{abstract}
\pacs{ }
\maketitle

\section{\label{intr}Introduction}

Observations in the last decade have gradually been providing evidence suggesting that not only most of the universe's energy is in a dark, unknown form, but indeed the dominant component of this dark sector violates the strong energy condition---the latter being required to explain the recent acceleration of the Universe. From a purely phenomenological point of view, the simplest candidate for this task is Einstein's cosmological constant (leading to the so-called $\Lambda$CDM or concordance model), and indeed this is in decent agreement with most of the existing data. However, from a fundamental physics point of view, the vacuum energy density suggested by observations is many orders of magnitude smaller than the most optimistic expectations, so that models where dark energy is dynamical (for example, being due to a scalar field) are arguably the more likely ones.

While there is, in some sense, a single constant dark energy model, the number of possible dynamical dark energy models is clearly infinite. Moreover, the standard observational techniques in use for probing dark energy are not ideal, in the sense that they do not measure directly the theoretically relevant quantities, and they allow for several important degeneracies to remain when comparing theory and observation. Both of these mean that even in an era of precision cosmology there is ample room for building phenomenological models of increasing complexity, whose differences when it comes to cosmological observables can be quite small, and hence are not easy to distinguish even with good data.

One may legitimately ask if these are merely `epicycles' which are doomed to be replaced by something entirely different and much simpler. While a fairly strong case can legitimately be made for a positive answer to this question, what should replace them still remains to be determined. What one can certainly say at this stage is that increasingly elaborate mathematics should not be used as an excuse to obscure simple physics. In other words, if one builds more elaborate models on the basis of pre-existing paradigms, then the broad general features of the original paradigm will still be present and can seldom be avoided, even if there are differences in the specific details.

The goal of the present paper (which is a follow-up and in the same spirit of \cite{Avelino3}) is to present a physically clear discussion of the clustering properties of dynamical dark energy models, specifically those where the dark energy is due to a classical scalar field with a perfect fluid form. We shall be relating our discussion to other recent literature, contrasting the different approaches in order to clarify crucial issues. We will provide an explicit simple proof that in quintessence-type models the dark energy fluctuations, on scales smaller than the Hubble radius, are of the order of the perturbations to the  Newtonian gravitational potential, and hence necessarily small on cosmological scales. We show that the often used homogeneous approximation is unrealistic and does not provide useful information, at least on sub-horizon scales. Moreover, we illustrate two physically simple points that are often obscured by unnecessary mathematics: the fact that fluctuations comparable to the Newtonian gravitational potential can be obtained by a mere gauge transformation, and that 
 the so-called dark energy mutation is a trivial artifact of an effective, single fluid description. Finally, we discuss the particular case where there is a coupling between the dark energy and dark matter fluids. Throughout this paper we shall work in natural units and a metric signature ($-,+,+,+$).

\section{\label{scalar}Scalar fields}

We shall be interested in the action
\begin{equation}\label{eq:L}
S=\int d^4x \, \sqrt{-g} \mathcal \, {\cal L}(X,\phi) \, ,
\end{equation}
where $\mathcal L$ is the Lagrangian for a real scalar field $\phi$ and
\begin{equation}\label{eq:kinetic_scalar1}
X=-\frac{1}{2}\nabla^\mu \phi \nabla_\mu \phi \,,
\end{equation}
is the kinetic term. We shall generically consider the case of a classical scalar field $\phi$ governed by the an arbitrary Lagrangian of the form ${\mathcal L}(X,\phi)$. We can write the energy-momentum tensor for this model in a perfect fluid form
\begin{equation}\label{eq:fluid}
T^{\mu\nu} = (\varepsilon + p) u^\mu u^\nu + p g^{\mu\nu} \,,
\end{equation}
by means of the following identifications
\begin{equation}\label{eq:new_identifications}
u_\mu = \frac{\nabla_\mu \phi}{\sqrt{2X}} \,,  \quad \varepsilon = 2 X p_{,X} - p \, ,\quad p =  {\mathcal L}(X,\phi)\, .
\end{equation}
In Eq.~(\ref {eq:fluid}), $u^\mu$ is the 4-velocity field describing the motion of the fluid (for timelike $\nabla_\mu \phi$), while $\varepsilon$ and $p$ are its proper energy density and pressure, respectively. Observe that from this it trivially follows that if $p=p(X)$, then $\varepsilon = \varepsilon(X)$. Unfortunately it is not always possible to invert $\varepsilon(X)$ and  obtain $X(\varepsilon)$, but when it is the fluid has an \emph{explicit} isentropic equation of state $p=p(\varepsilon)$.

In the special case where $\mathcal L (X,\phi) = f(X)-V(\phi)$ (with $f$ and $V$ being arbitrary functions of $X$ and $\phi$, respectively), the mass of the scalar field is defined as
\begin{equation}
m^2 \equiv \frac{\partial^2 V}{\partial \phi^2}\,.
\end{equation}
For a viable dark energy model one typically requires a mass smaller than the Hubble parameter $H$, $m \lsim  H$, at least for a canonical scalar field with $f(X)=X$. On small (effectively sub-horizon, $\lsim m^{-1}$) scales, pressure and density fluctuations are related through $\delta p = c_s^2 \delta \rho$ where the sound speed is given in linear theory by \cite{Mukhanov}
\begin{equation}
c_s^2 = \frac{p_{,X}}{\varepsilon_{,X}}=\frac{{\mathcal L}_{,X}}{{\mathcal L}_{,X}+2X{\mathcal L}_{,XX}}\,. \label{soundspeed}
\end{equation}
If $f(X)=X$ we have a canonical scalar field which has a constant sound speed of unity. However, we have the freedom to choose the function $f$ in order to obtain very different sound speeds. An example of an algebraically simple but physically interesting class of functions is $f(X) \propto X^n$, which yields $c_s^2= 1/(2n-1)$. In particular, when $n=1$, the scalar corresponds to a massless scalar field, $n=2$, to background radiation, and so on. In the limit $n \rightarrow \infty$, the scalar can be interpreted as dust (in other words, pressureless non-relativistic matter).

On the other hand, the scalar field equation of state is defined by
\begin{equation}
w \equiv \frac{p}{\varepsilon}=\frac{\mathcal L}{2 X {\mathcal L}_{,X} - {\mathcal L}}=\frac{f(X)-V(\phi)}{2 X f_{,X} - f +  V(\phi)}\,. \label{eos}
\end{equation}
Observe that by carefully choosing $V(\phi)$ we may independently specify $w$ and $c_s^2$. This simple point will be important for some of what follows. 
Finally, we can also differentiate Eq.~(\ref{eos}) and substitute Eq.~(\ref{soundspeed}) to find
\begin{equation}
w_{,X} = (c_s^2-w)\frac{\varepsilon_{,X}}{\varepsilon}\,, \label{running}
\end{equation}
which shows that $w=c_s^2$ is sufficient to ensure that the equation of state has no dependence on $X$. It is not, however, necessary as it also occurs whenever the energy density itself has no such dependence ($\varepsilon_{,X}=0$); the cosmological constant is a simple example.

\section{Homogeneous approximation}

Many aspects of the evolution of dark energy perturbations have been studied in great detail, particularly in the context of standard quintessence scenarios. These are characterized by two main components, dark matter (DM) and dark energy (DE), and the latter is usually modeled as a canonical scalar field with $f(X)=X$. As shown in the previous section, dark energy in these models is characterized by $c_s^2=1$ (and $w \sim -1$). However, in many of these studies, only the homogeneous case has been investigated \cite{Nunes,Abramo2,New2}. For our present purposes, the key feature of this approximation is that pressure gradients are completely absent.

Although the homogeneous approximation greatly simplifies the analysis, it is not difficult to see, even in the absence of any detailed comparisons, that the corresponding results are unrealistic. Indeed, such an approximation can only be considered reliable when uniform perturbations on scales larger than the Hubble radius are considered, which is clearly not the situation of greater physical interest. Still, before proceeding to our main analysis, we will briefly review the homogeneous case (in which dark matter and dark energy collapse together), so as to provide a useful comparison point.

Let us assume that the energy-momentum tensors of DE and DM are separately conserved, in other words, that there is no coupling between them (we will discuss the coupled case later in this note). If we consider the evolution of a region of physical volume $\mathcal V$, we can write
\begin{equation}
{d \varepsilon}_{\rm DE} + (1+w)\varepsilon_{\rm DE} \frac{d {\mathcal V}}{\mathcal V} =0 \,
\end{equation}
\begin{equation}
{d \varepsilon}_{\rm DM} + \varepsilon_{\rm DM} \frac{d {\mathcal V}}{\mathcal V}=0 \, ,
\end{equation}
so that
\begin{equation}
\delta_{\rm DE}=(1+w) \delta_{\rm DM}\, ,
\end{equation}
where $w = p_{\rm DE}/\varepsilon_{\rm DE}$ is the dark energy equation of state, $\delta=(\varepsilon-{\bar \varepsilon})/{\bar \varepsilon}$ is the contrast and ${\bar \varepsilon}$ denotes the average density.

Simple as this is, it is sufficient to highlight an important consequence of this approximation. For generic values of the equation of state parameter $w$ (\emph{viz.}, values not incredibly close to $w=-1$), large fluctuations in the dark matter component will inevitably be accompanied by significant fluctuations in the dark energy component. As we will see in the next section, this is in fact not the case for realistic sub-horizon perturbations.

\section{Linear fluctuations}

An order-of-magnitude estimate of the perturbations in the dark energy component can be obtained by considering linear theory. We will therefore consider a standard dark matter fluid with $w_{\rm DM}=0$ and a dark energy fluid with and arbitrary equation of state $w$ and sound speed $c_s^2$. Unless otherwise stated we shall be working in the synchronous gauge, which is comoving with the dark matter, that is $u^i_{\rm DM}=0$.

The linear evolution of scalar perturbations is described by
\begin{equation}\label{eq:pertub_1}
\ddot \delta_{\rm DM}+{\cal H}\dot\delta_{\rm DM}-\frac{3}{2}{\cal H}^2\left[\Omega_{\rm DM}\delta_{\rm DM}+(1+3c_s^2)\Omega_{\rm DE}\delta_{\rm DE}\right]=0 \,
\end{equation}
\begin{equation}\label{eq:pertub_2}
\dot \delta_{\rm DE} + 3 {\cal H}(c_s^2 - w) \delta_{\rm DE}  + (1+w) (\theta - \dot \delta_{\rm DM})= 0 \,
\end{equation}
\begin{equation}\label{eq:pertub_3}
\dot \theta + {\cal H}(1 - 3 c_s^2) \theta + \frac{c_s^2}{1+w} \nabla^2 \delta_{\rm DE}= 0 \, 
\end{equation}
where a dot denotes an  $\eta$ conformal time derivative (with $d \eta = dt /a$ and ${\cal H}=a H={\dot a}/a$), $\Omega_{\rm i}= \varepsilon_{\rm i}/\varepsilon_{\rm c}$ ($\varepsilon_c=3 H^2/8\pi G$ being the critical density), and we have also defined
\begin{equation}\label{deftheta}
\theta=\nabla \cdot {\vec v}_{\rm DE}=a(u^i_{\rm DE})_{,i}\,.
\end{equation}

On scales much smaller than the Hubble radius (that is $k {\cal H} \gg 1$), and assuming for simplicity that the dark matter fluctuations are the only source for the dark energy fluctuations, we can write
\begin{equation}
 \frac{c_s^{\, 2}}{1+w} \nabla^2 \delta_{\rm DE} \sim \frac{3}{2}{\cal H}^2 \Omega_{\rm DM}\delta_{\rm DM}=4\pi G a^2 \delta \epsilon_{\rm DM}\,.
\end{equation}
If we now consider a density perturbation of comoving wave number $k$ one immediately finds that
\begin{equation}
\frac{c_s^2 k^2 \delta_{\rm DE}}{1+w} \sim 4\pi G a^2 \delta \varepsilon_{\rm DM} \,.
\end{equation}
Equivalently, we can write this as a function of the characteristic scale of the perturbation $L \sim a/k$,
\begin{equation}
\frac{\delta_{\rm DE}}{1+w} \sim \frac{GM}{L} \,,
\end{equation}
where we have taken the sound speed $c_s^2=1$ and also defined $M = \delta \varepsilon_{\rm DM} L^3$. This shows that any fluctuations in the dark energy component are of the order of the perturbations to the Newtonian gravitational potential and thus necessarily very small on cosmological scales. This result agrees with those of \cite{Mota,New1}, but has been obtained in a physically clearer way.

Indeed, with hindsight this result is hardly surprising. A simple way to understand it is to observe that in the standard quintessence scenario, in order to have $w \sim -1$, one requires $|X/V| \ll 1$. It then follows that the amplitude of the fluctuations on sub-horizon scales must necessarily be small in this type of models. For this not to be the case, the kinetic term $X$ would wave to be substantial and we would no longer have a dark energy fluid. The only way to get around this restriction is to consider other classes of dynamical dark energy models.

However, this is not yet the full story. It must be emphasized that fluctuations of this same order are also associated with different gauge choices. For example, by considering a gauge transformation from a local inertial frame comoving with the dark matter fluid (which we will denote as frame I) to the local inertial cosmological frame (denoted by frame II, with a vanishing CMB dipole), one obtains 
\begin{equation}
\Delta\equiv\frac{\varepsilon_{\rm DE,II}-\varepsilon_{\rm DE,I}}{\varepsilon_{\rm DE,I}}=\gamma_D^2-1+w \, v_D^2 \,
\end{equation}
and this can be approximately written
\begin{equation}
\Delta\sim (1+w)v_D^2 \sim (1+w) \frac{GM}{L} \,,
\end{equation}
where $v_D$ is the velocity of the dark matter fluid with respect to frame II and $\gamma_D=(1-v_D^2)^{-1/2}$. In the last step we have made the reasonable assumption that the main contribution for the dipole is directly related to the local Newtonian gravitational potential induced by a mass perturbation $M$ with length scale $L$. Perturbations of comparable magnitude may be similarly obtained by considering other possible gauge choices.

\section{Dark energy mutation}

The analysis of the previous section shows that if the sound speed of the dark energy fluid is significant, then the dark energy perturbations must necessarily be small. However, for very small values ($c_s^2 \sim 0$) the above statement is no longer true, so this case warrants a separate treatment. In this section we shall therefore consider a model with a null sound speed $c_s^2=0$, by requiring the pressure $p$ to be constant, and study the behavior of a region of physical volume $\mathcal V$.

Energy-momentum conservation trivially implies 
\begin{equation}
\label{eq2a}
d \left[(\varepsilon+p){\mathcal V}\right] = 0 \,,
\end{equation}
and consequently $(\varepsilon+p)\propto {\mathcal V}^{-1}$. The equation of state can be written in the form
\begin{equation}
\label{eq2b}
w=\frac{p}{\varepsilon}=\frac{1}{C {\mathcal V}^{-1}-1} \, ,
\end{equation}
where $C$ is a constant. This simple calculation shows that $w$ will depend on the physical volume $\mathcal V$ of the region under consideration and, consequently, on the density perturbations. This effect is just a coarse-graining issue, in the sense commonly discussed in condensed matter systems. We see that in low-density regions, where $\mathcal V$ is very large, $w \sim -1$, but in collapsed regions the value of $w$ can be much smaller (in modulus) and even approach zero: indeed, in the limit ${\mathcal V} \to 0$, we have $w \to 0$. 

The most noticeable consequence emerging from this analysis is that if density perturbations are present, then it necessarily follows that $w$ cannot be a constant. This effect has been pointed out in \cite{Abramo1} and dubbed `dark energy mutation'. However, there is nothing surprising about it, and indeed its physical explanation is rather prosaic. This model turns out to be the simplest example of unified dark energy models studied in \cite{Avelino2,Beca,Avelino3}, and it has been shown to be totally equivalent (to any order) to $\Lambda$CDM. In fact, there is no way to distinguish the single fluid interpretation made above from the standard interpretation with  canonical components, dark matter and a cosmological constant, having respectively $w_{\rm DM}=0$ and $w_{\rm DE}=-1$. Dark energy mutation, in this case, is just an artifact of our single fluid description.

\section{Coupled models}

In the so-called concordance model, a range of observational data is used to postulate the existence of two dark fluids (DE and DM) for which so far there is no direct experimental backing. In the context of GR, the most common attitude is to model DE and DM as two minimally coupled fluids. The direct opposite to this is to treat them as different manifestations of a single fluid (UDE models). An intermediate approach, on the other hand, is to view them as coupled fluids. In this case, however, if the coupling is very strong, we naturally expect the distinction of DE and DM as two different fluids to become somewhat blurred. In other words, we expect that, to a certain extent, strongly coupled fluids will behave as if a single fluid. As far as we are aware today, this bridge between strongly coupled models and UDE has not been significantly explored.

Following the recent literature \cite{Beca,Brax,Afshordi,Kaplinghat,Bean1,Bean2,Avelino3}, we shall assume that the two dark fluids are coupled through the Lagrangian
\begin{equation}\label{lagra_1}
{\mathcal L}=X-V(\phi)+h(\phi){\mathcal L}_{\rm DM}\,.
\end{equation}
($\phi$ in this context is normally called a `chameleon' field \cite{New3a,New3b,New3c,Brax}.) Note that according to the discussion in Sect. \ref{scalar}, the dark matter component may be described by a scalar field $\varphi$ with a Lagrangian
\begin{equation}
{\mathcal L}_{\rm DM} \propto Y^n\,,
\end{equation}
in the limit of large $n$, where
\begin{equation}\label{eq:kinetic_scalar2}
Y=-\frac{1}{2}\nabla^\mu \varphi \nabla_\mu \varphi \, .
\end{equation} 
 Indeed, for a fixed value of $n$, $\varphi$ has an equation of state parameter $w_{\rm DM}=(2n-1)^{-1}$ and, therefore,  $p_{DM}=w_{\rm DM} \varepsilon_{\rm DM}$ becomes negligible for large $n$. If follows that we can rewrite (\ref{lagra_1}) as
\begin{equation}
{\mathcal L}=X-V(\phi)+g(\phi)\varepsilon_{\rm DM}
\end{equation}
where  $g(\phi)=w_{\rm DM} h(\phi)$ is a rescaled coupling constant. It is easy to check (by varying the action in relation to $\varphi$) that the dark matter component evolves independently from the chameleon field. On the other hand, the evolution of $\phi$ is given by
\begin{equation}\label{motion}
\square \phi=\frac{\partial V_{\rm eff}}{\partial \phi} \, ,
\end{equation}
where
\begin{equation} \label{v_effective}
V_{\rm eff}=V(\phi)-g(\phi)\varepsilon_{\rm DM}\,.
\end{equation}
and therefore is affected by how dark matter evolves. (Note here that although $V_{\rm eff} \simeq V$, $\partial V_{\rm eff} / \partial \phi$ can be very different from $\partial V/ \partial \phi$.) As for the energy-momentum tensor associated with (\ref{lagra_1}), it is a simple matter to show (by varying the action in relation to $g^{\mu\nu}$) that
\begin{eqnarray}\label{T_chameleon}
T_{\mu\nu} (\phi,\varphi)&=& \nabla_\mu \phi \nabla_\nu \phi 
+  (X-V(\phi))g_{\mu\nu} + \nonumber \\ 
&+ & h(\phi) \left[g_{\mu\nu} Y^n + n Y^{n-1} \nabla_\mu \varphi \nabla_\nu \varphi \right] \,. 
\end{eqnarray}

Obviously, this energy-momentum tensor does not, in general, describe a perfect fluid. However, in the so-called \emph{adiabatic regime} (described, in detail, in \cite{Bean1,Bean2}), it is assumed that the gradients of $\phi$ are both negligible in $T_{\mu\nu}$ and the equation of motion (\ref{motion}). Thus, in this regime, \label{T_chameleon} reduces to
\begin{equation}
T_{\mu\nu} \simeq (hY^n-V)g_{\mu\nu} +  n hY^{n-1} \nabla_\mu \varphi \nabla_\nu \varphi  \,, 
\end{equation}
which can be immediately written in a perfect fluid form, if we make the following definitions
\begin{eqnarray}\label{identifications}
u_\mu &=& \frac{\nabla_\mu \varphi}{\sqrt{2Y}} \,,  \quad \varepsilon_{\rm eff} = h \varepsilon_{\rm DM} + V \, ,\nonumber \\ 
p_{\rm eff} &=&  -V_{\rm eff} \simeq -V \, .
\end{eqnarray}
The effective equation of state is then
\begin{equation}
w_{\rm eff} = \frac{p_{\rm eff}}{\epsilon_{\rm eff}} \simeq \frac{-V(\phi)}{V(\phi) + h \varepsilon_{\rm DM}} \, .
\end{equation}
Is it an isentropic fluid, though? Yes. Since the adiabatic regime is also characterized by the condition $\partial V_{\rm eff}/ \partial \phi = 0$, the value of $\phi$ is univocally related to $\varepsilon_{\rm eff}$. Hence, $p_{\rm eff}$ only depends on the value of $\varepsilon_{\rm eff}$ and therefore the fluid is isentropic (although, in general, we won't have an explicit $p_{\rm eff} = p_{\rm eff}(\varepsilon_{\rm eff})$ equation of state).

Now, the value of
\begin{equation}
m^2_{\rm eff}\equiv \frac{\partial^2 V_{\rm eff}}{\partial \phi^2} \, ,
\end{equation}
determines the length  scales for which the adiabatic regime is valid. Specifically, this is the case for large scale perturbations with $L \gg m_{\rm eff}^{-1}$, while for scales much smaller that this, the approximation is no longer valid. We thus conclude that above a certain scale, sufficiently coupled models behave as a single isentropic  fluid but not below. Why is this relevant? It is relevant because the background evolution in UDE models described by a single isentropic fluid is expected to be strongly affected by non-linear effects which severely complicate the analysis of this type of models. On the other hand, it is still unclear if the differentiated behaviour above or below $L \sim m_{\rm eff}^{-1}$ may help to relax the averaging problem \cite{Beca} that affects UDE models. For the sake of argument, suppose the majority of the non-linear clustering occurs for scales smaller that $m_{\rm eff}^{-1}$; since now they are confined to a non-isentropic part of the fluid, it is possible that they may not greatly affect the large scale evolution of the universe. If, on the other hand, significant clustering does extend beyond this scale, then non-linearities will still be a major problem in strongly coupled models (at least for models not sufficiently close to a $\Lambda$CDM model).

Finally, note that at recent epochs, 
\begin{equation}
|V(\phi)| \gg | g(\phi) \varepsilon_{\rm DM}| \, ,
\end{equation}
and, in the adiabatic regime
\begin{equation}
V^\prime(\phi) - g^\prime(\phi) \varepsilon_{\rm DM} \simeq 0 \, ,
\end{equation}
where ${}^\prime$ denotes $\partial/\partial \phi$. It follows that
\begin{equation}
\left|\frac{V^\prime(\phi)}{V(\phi)}\right | \ll \left| \frac{g^\prime(\phi)}{g(\phi)} \right | \simeq \left| \frac{G^\prime(\phi)}{G(\phi)}\right | \, .
\end{equation}
Given the stringent astrophysical and laboratory constraints on variations of Newton's constant $G$,  the above relation imposes strong constraints on the shape of the potential in the region explored by the field in recent times. 
 
\section{\label{conc}Conclusions}
In this paper we studied in some detail the clustering properties of dynamical dark energy models. We have shown that in standard quintessence models, contrary to recent claims, dark energy fluctuations on sub-horizon scales are of the order of the perturbations to the Newtonian gravitational potential and, consequently, very small on cosmological scales. We have also pointed out that the homogeneous approximation is not an adequate framework to study the evolution of dark energy fluctuations on small scales.

We have also explored the extent to which coupled models can be interpreted as unified models. Our analysis makes it clear that the non-linear instability that plagues unified dark energy models \cite{Avelino3} should also apply to these so-called chameleon models. A more detailed analysis of this scenario is warranted, but we leave it for future work.

\bibliography{UDE3}
\end{document}